# The Rise of a New Digital Third Space Professional in Higher Education: Recognising Research Software Engineering

Shoaib Sufi, University of Manchester

## Introduction

Research Software Engineering is the application of professional software skills to research problems. Those who do this are called Research Software Engineers or RSEs for short. RSEs work closely with researchers in a collaborative fashion rather than just offering a standalone function (c.f. the traditional IT workforce or Librarians working to provide a general set of services to the University community such as research, teaching or administrative functions). It is this overlap with academic researchers that make the RSE's and the RSE management a new type of third space professional in the higher education sector.

In Reconstructing Identities in Higher Education (RIHE) (1) Whitchurch highlights the Third Space as a sphere between professional and academic spheres in Higher Education (HE) involving interactions between teams and networks working in parallel with normal University structures that give rise to new forms of management and leadership. It is by its nature a grey area, one which has a greater freedom for experimentation and creativity but also a more uncertain place in terms of role in the University.

The roles identified by Whitchurch include those supporting student life, widening participation, learning support, community partnerships, research and business partnerships and institutional research on page 28 of RIHE (1). RIHE could not have foreseen the rise of RSE's and its data equivalent (Data Stewards). The term Research Software Engineer (RSE) was coined in 2012 (2) after initial discussions at the UK Software Sustainability Institutes annual Collaborations Workshop[1].

We explore the rise of the RSE role and how and this maps onto Third Space professionalism and why this is important and what it holds for the future of the Third Space and perhaps ways in which HE can better cater for the needs of these new classes of digital professionals.

---

[1] https://www.software.ac.uk/programmes-and-events/collaborations-workshops

# The rise of the Research Software Engineer

The Research Software Engineering movement arose out of a recognition of academic software developers being largely overlooked and having no proper career structures (2). From initial discussions at the UK Software Sustainability Institute's 2012 Collaborations Workshop[2] members of the original group went on to write a piece for Digital Research 2012 (3) fleshing out their key thoughts.

They highlighted the increased use of the digital in research, the lack of recognising professional best practice in research software and the skills needed to maintain research software. They touched upon matters of reproducible research and the need for a new role to be recognised - one that recognised individuals who were skilled in research and who primarily did professional software development. Hence the term 'Research Software Engineer' was coined.

The fact that many RSEs have skills such as training, writing papers, contributing to grant proposals, writing technical reports and manuals as well as keeping up to date with IT trends in numerical algorithms, computer languages and analysis tools was also highlighted. In a sense RSEs really did bridge the software to research function and were thus very much in the Third Space as highlighted by RIHE by sitting between academic practice and pure administrative practice. Career development (or lack thereof was highlighted); another aspect common to the Third Space.

What was not originally envisaged was the need for RSE groups and RSE management, functions which naturally sat as branches of the central IT function but given lots of leeway to do things as they saw fit. A community effort was needed to develop the fledgling functions at the different institutions and this led to the creation of the RSE leaders network[3] where former RSEs who were now trying to create RSE groups at institutions could network to share best practice.

In addition to this the funding agencies (in the form of UKRI EPSRC) were keen on the idea of the RSE and funded specific Fellowships to help establish institutional and domain groups to improve software practices in those areas There were calls in 2015 and 2017 providing 5 years of funding to help bring about change[4].

As the community grew there was a greater need to network and share experiences this led to the first RSE Conference in 2016[5] and then annually since with ever increasing number of attendees leading to a need to increase the capacity of the venue (during the pandemic of 2020/21 it was replaced by the SORSE[6] initiative of online events).

---

[2] https://www.software.ac.uk/cw12
[3] https://society-rse.org/community/rse-leaders-meetings/
[4] https://society-rse.org/community/rse-fellows/
[5] https://society-rse.org/events/
[6] https://sorse.github.io/

Almost 30 Institutions[7] now have RSE groups and the associated RSE management and organisational buy in, however given that there are 165 Higher Education Institutions in the UK[8] there is some way to go before this set of roles is ubiquitous.

A Research Software Engineering Association[9] led the organisation of RSE events in the UK until March 2019 when a formal charity was registered, 'The Society of Research Software Engineers'[10] to act as a vehicle to bring together members of the community and run the annual events and act as a single focus group to promote the practice of Research Software Engineering to institutions, funders and policy makers.

The wider RSE movement is about making the implicit and invisible work that was happening supporting research software more explicit, visible and valued.

## Mapping RSE roles onto the Third Space

The number one characteristic of Third Space identities is the project based roles their work entails. There is a strong emphasis on partnership with academic colleagues taking the place of the traditional service to academic colleagues. Although Third Space roles are 'under the radar' it's this exact problem that the naming, formation and growth of the RSE community hopes to alleviate. It's clear that management of Third Space staff by those working in the purely professional or academic spheres needs informing to help them understand the unique set of circumstances they work under and how to facilitate them in line with their role. To this end RSE groups and RSE management structures are the working unit at institutions, however there are still embedded RSEs who work attached to larger research groups that are led by academics[11]

In the following sections we explore notions of knowledge, relationships, legitimacies and languages for the RSE role in the context of their definition in RIHE to help us identify where the RSE roles (both the engineers themselves and the management of the units) resonate and map onto facets of the Third Space.

After this we briefly discuss the type of HE professional RSEs are, exploring whether they class as 'unbounded' or 'blended'.

### Knowledge

RSEs are knowledge brokers, they have a wide range of software development and analysis skills that they can offer to different disciplinary areas. The knowledge RSEs have and their managers tends to be propositional (4), i.e. it is adapted and applied to specific circumstances and actions rather than being about rolling out as a stable process.

---

[7] https://society-rse.org/community/rse-groups/
[8] https://www.universitiesuk.ac.uk/facts-and-stats/Pages/higher-education-data.aspx
[9] https://rse.ac.uk/about/
[10] https://register-of-charities.charitycommission.gov.uk/charity-search/-/charity-details/5125299/governance
[11] Such as the eScience Lab (https://esciencelab.org.uk/) and the Health eResearch Centre (https://www.herc.ac.uk/) both at the University of Manchester.

The assignments RSEs undertake are bespoke, it's not just about what an RSE might know but about the ability to pick up new information and adapt their approach to the problem at hand.

## Relationships

Third space professionals have strong people and client orientation (1). This is what RSEs and their manager do, they work with academic researchers thus their technical knowledge and communication skills have to be at the right level to make this relationship work.

While the relationships with academics is key, bridging the possibility and expectation gap is an important aspect of the work. Academics are not always versed in the possibilities of software and through conversation and trust the topics such as what academics think may be hard to do but they turn out to be easy and those that the academics think are easy to do and turn out to be impossible can be broached and agreement can be settled on[12]. As mutual respect is built up, the academics may even invite the RSE to participate in working on software aspects of a funding proposal.

RSE managers also have an outward focus and they have projects (along with understanding academics) where they are trying to build stronger links with industry. A case in point is the Institute of Coding[13] which is pushing for working with industry to produce software curricula which will lead to greater preparedness for people to work in industry. This blurring of the roles between academics, teaching and learning and professional projects and reaching outside the institution to work with industry is very much a feature of Third Space professionals.

## Legitimacies

Authority in Third Space roles is self constructed (1). This may apply more to the RSE function as a whole (it's management, leadership and individual RSEs). If the RSE does a good job and works well with the academics and the RSE manager has a professional relationship with the academic and manages expectations and the RSE unit delivers value for the academic then there is a greater chance that the RSE staff will get repeat business and an improved reputation. There are no organisational structures mandating that the academic work with the RSE unit, the academics work with them because they are convinced they will be helped - not because they have to.

In terms of credentialled legitimacy, while a PhD is not needed for the RSE post they certainly don't harm in understanding the research process and having been through the academic training it can be something which gains more credibility from the academics being worked. Many RSEs do have research training, for many it's what leads them to wanting to become an RSE, they want to focus more on the software rather than on having their own research agenda.

---

[12] Taken from conversation with an RSE Group leader at a Russell Group University.
[13] https://instituteofcoding.org/

Results from the International RSE survey 2018 (5) indicate over 50% of the RSEs who answered the survey had doctorate level training. Ultimately a record of delivery is what counts.

Working on different briefs with different technologies, domains and problems is a win for the RSE who hopes to gain varied experience. The academics gain from this by knowing that the RSE will bring varied experience to their specific problems. The RSE also avoids being pigeonholed into one particular type of task.

RSE's and the academic's can start off as a knotworking (6) arrangement - coming together on a temporary basis for a specific project or purpose and with an aim to really accelerate progress but this can form a mutual and more long running engagement if this is mutual beneficial to the RSE, RSE Management and the academic.

## Language

Whitchurch in (1) analysed the use of language used for recruiting Third Space professionals. There is very formal language used around contractual obligations and structured working around the maintenance of systems along with controlling words such as 'ensure'. There is also the use of more fluid language around project responsibilities such as 'creativity' and 'innovation'. Alongside this there are active verbs used such as 'developing','analysing' and interpreting'.

What is fascinating and affirming of the RSEs identification in the Third Space is the extent to which this aligns to the role that RSEs perform. RSEs are working often in projects which have this notion of contractual obligations to meet specified deliverables. The software projects are not only experimental but they also have notions of infrastructure provision which merge with the idea of maintenance and service level agreements for production grade aspects of the projects. RSEs are recruited for their creative problem solving and innovative solutions to research software problems. All of this work involves constructing systems, analysing data and offering algorithms that offer varied interpretation to give the academic a better choice of possible solutions.

The RSE's know they cannot be 'geeks in the corner' so have to be able to interact and bridge between the academics. RSE managers do choose specific people for specific roles as they have a background in the research domain or have a track record of being able to bridge. This also points to the need and practice of different levels of RSE from junior through to senior RSEs. With senior roles expected to need minimal coaching when it comes to effective communication in large scale projects.

The language and facilities of RSE units are not fully integrated yet into the working practices of Research in HE as not all institutions have them[14]. The name Research Software Engineer is also important; it gives credit to the fact that research is important, just calling them Software Engineers would not be as comforting for the academics - RSE was an intentional terminological choice (3). RSE roles by definition want to understand the research process they are helping and it's in the academics interest to explain what is happening and research aims - this harmony helps to form 'mediating artefacts' - a common language (7) to help move towards a common goal.

---

[14] https://society-rse.org/community/rse-groups/

Shared terminology can work in both directions with academics informing RSEs of necessary terminology they may not be aware of and the RSE characterising computational aspects which could benefit the academic to think in terms of, e.g. 'automation' of repetitive task, being able to handle more data - 'scale', being able to automate different facet and facet combinations in an analysis - 'a parameter sweep', to better engineer software to make it more resilient to failure - 'hardening of codes' and the idea of 'well managed software' (making it easier to reuse and build upon in the future).

### Moving from unbounded to blended professionals

RSE units at institutions tend to be formed of what was referred to in (1) as 'blended-professionals', those who are specifically recruited to work across the professional and academic domains.

Many RSE are coming from a research background or working on research projects. We mentioned embedded RSE units where they are focused on the projects of one group; their mindset with regards to the organisation tends to be more as unbounded-professionals - they are focused on their needs of their projects rather than any organisational perspective. These embedded RSE are not so much of an aberration but are very much an original form of RSE before the term was created and their are valid reasons for them still to exist such as taking a greater responsibility managing the relationships, requirements gathering and consensus building in large research projects; this leaves very little time for interacting with their home institutions.

## What the future holds

Research Software Engineering, as an area is very much a work in progress; there are mature groups and now a Society but with 30 RSE units and over 165 institutions of Higher Education (in the UK) there is some way to go to make the structures and base offering of the RSE movement ubiquitous. The very clear role of RSE as working with academics (but being aligned in the organisation as an offshoot of the central IT services function) makes them very deliberate Third Space professionals.

There are many ways in which the RSE could contribute to the wider organisations that they operate in:

- Contributing to curricula development e.g. in computational research and in applied computer science as they know what works.
- The RSE can be seen as a parallel track by academics who might switch to if they want to create a software library of methods for a particular scientific domain e.g. via an RSE Fellowships; perhaps there could be a mixed careers with an easy movement between the two roles.
- The realisation of benefit through centralisation of RSE provision to bring together the knowledge and experience on software across the University

- Practice can be ahead of procedures so the RSE units can have a view of which organisational processes are useful and which need to become more efficient; this RSE units should be consulted in organisational improvement projects (an example of this is the Research Life Cycle Programme[15] at the University of Manchester which has consulted those in RSE and other Third Space roles).

There are some open problems though, and one of the main ones is around career paths. Various roles have appeared in RSE units from Junior RSE, RSE, Senior RSE to the more management and leadership focused roles. However a purely technical track involving business analysis skills, software architecture, coding and analysis skills is yet to exist in the RSE units, this leads to a natural ceiling on where technical skills can lead a person. Perhaps motivation and a template can be taken from the IBM Fellows[16] track which are supported as very senior technical staff (naturally they are somewhat outstanding!). Motivations, incentives and rewards are a topic for another analysis on RSEs, however it is an important area to make progress on. Ultimately it is beneficial for research in the long term to stem the loss of expertise and lowering staff turnover. It's an open problem as to how we make those in these roles less 'invisible' and less 'imposter' (8). RSE have parallels with technicians, perhaps institutions will sign up to a future RSE commitment in the same way they have signed up to the Technicians Commitment[17].

## Conclusion

RSEs roles, whether embedded, or organisational units are operating across the academic and professional services divide as well as being very project focused roles. RSE is thus a Third Space role.

The knowledge they are gaining, the relationships they are formulating by working with academics and those outside the institution, the legitimacies they are creating by doing good work and gaining the trust of the academics and the language that is used to construct their roles as well as what they formulate and understand from their varied assignments all point to the fact that RSE are Third Space.

RSE units are not ubiquitous in HE, around 20% of HE institutes (in the UK) have them so there is some work needed to promote this way of working. This win-win way of working that supports academics essentially trying out interacting with research software specialists creates a safe space for innovation and creativity; all hallmarks of the Third Space. If academics choose to work with the RSE, they are not responsible for their career development or contracts, just paying for a portion of their time from their research grants making it far less onerous on academics to get expert software advice.

---

[15] https://www.rlp.manchester.ac.uk/
[16] https://www.ibm.com/ibm/ideasfromibm/us/ibm_fellows/
[17] https://sciencecouncil.org/employers/technician-commitment/

Ultimately Research Software Engineering being Third Space legitimises the need for better career development and closer organisational management with other Third Space professionals. The Research Software Engineering function makes research better by improving the integrity and reproducibility research through better software, this should be championed by academics to raise the issue of nurturing and rewarding RSEs to an organisational level.

# References

1. Reconstructing Identities in Higher Education: The rise of "Third Space" professionals [Internet]. Routledge & CRC Press. [cited 2020 Dec 22]. Available from: https://www.routledge.com/Reconstructing-Identities-in-Higher-Education-The-rise-of-Third-Space/Whitchurch/p/book/9780415614832
2. A not-so-brief history of Research Software Engineers | Software Sustainability Institute [Internet]. [cited 2021 Jan 25]. Available from: https://www.software.ac.uk/blog/2016-08-17-not-so-brief-history-research-software-engineers-0
3. Baxter R, Hong NC, Gorissen D, Hetherington J, Todorov I. The Research Software Engineer. In Oxford; 2012. p. 4. Available from: https://www.research.ed.ac.uk/portal/en/publications/the-research-software-engineer(e8416ad7-750f-442f-9b17-d812b9bb414d)/export.html
4. Eraut M. Developing Professional Knowledge And Competence [Internet]. Routledge; 2002 [cited 2021 Jan 25]. Available from: https://www.taylorfrancis.com/books/developing-professional-knowledge-competence-michael-eraut/10.4324/9780203486016
5. Philippe O, Hammitzsch M, Janosch S, Anelda Van Der Walt, Werkhoven BV, Hettrick S, et al. softwaresaved/international-survey: Public release for 2018 results [Internet]. Zenodo; 2019 [cited 2021 Jan 25]. Available from: https://zenodo.org/record/1194668
6. Engeström Y. From Teams to Knots: Activity-Theoretical Studies of Collaboration and Learning at Work [Internet]. Cambridge: Cambridge University Press; 2008 [cited 2021 Jan 20]. (Learning in Doing: Social, Cognitive and Computational Perspectives). Available from: https://www.cambridge.org/core/books/from-teams-to-knots/E37D0858F603DE6A88281FA0D46BD856
7. Czarniawska B. On Creole Researchers; Hybrid Disciplines and Pidgin Writing. 2007 Nov 27 [cited 2021 Jan 25]; Available from: https://ep.liu.se/en/conference-article.aspx?series=ecp&issue=25&Article_No=8
8. Akerman K. Invisible imposter: identity in institutions. Perspectives: Policy and Practice in Higher Education [Internet]. 2020 Oct 1 [cited 2021 Jan 24];24(4):126–30. Available from: https://doi.org/10.1080/13603108.2020.1734683